\documentclass[aps,prl,floatfix,superscriptaddress,reprint]{revtex4-2}
\usepackage{mathrsfs,braket}
\usepackage{amssymb, amsbsy, amsmath, latexsym, dsfont, array, layout, graphicx, mathrsfs, color, ulem, bm}
\usepackage[colorlinks,linkcolor=blue,anchorcolor=red,citecolor=blue]{hyperref}

\begin{document}
\title{Anderson transition and mobility edges on hyperbolic lattices with randomly connected boundaries}
\author{Tianyu Li}
\affiliation{Beijing National Laboratory for Condensed Matter Physics, Institute of Physics, Chinese Academy of Sciences, Beijing 100190, China}
\author{Yi Peng}
\affiliation{International Quantum Academy, Shenzhen 518048, China}
\author{Yucheng Wang}
\email{wangyc3@sustech.edu.cn}
\affiliation{Shenzhen Institute for Quantum Science and Engineering, Southern University of Science and Technology, Shenzhen 518055, China}
\affiliation{International Quantum Academy, Shenzhen 518048, China}
\affiliation{Guangdong Provincial Key Laboratory of Quantum Science and Engineering, Southern University of Science and Technology, Shenzhen 518055, China}
\author{Haiping Hu}
\email{hhu@iphy.ac.cn}
\affiliation{Beijing National Laboratory for Condensed Matter Physics, Institute of Physics, Chinese Academy of Sciences, Beijing 100190, China}
\affiliation{School of Physical Sciences, University of Chinese Academy of Sciences, Beijing 100049, China}

\begin{abstract}
Hyperbolic lattices, formed by tessellating the hyperbolic plane with regular polygons, exhibit a diverse range of exotic physical phenomena beyond conventional Euclidean lattices. Here, we investigate the impact of disorder on hyperbolic lattices and reveal that the Anderson localization occurs at strong disorder strength, accompanied by the presence of mobility edges. Taking the hyperbolic $\{p,q\}=\{3,8\}$ and $\{p,q\}=\{4,8\}$ lattices as examples, we employ finite-size scaling of both spectral statistics and the inverse participation ratio to pinpoint the transition point and critical exponents. Our findings indicate that the transition points tend to increase with larger values of $\{p,q\}$ or curvature. In the limiting case of $\{\infty, q\}$, we further determine its Anderson transition using the cavity method, drawing parallels with the random regular graph. Our work lays the cornerstone for a comprehensive understanding of Anderson transition and mobility edges on hyperbolic lattices.
\end{abstract}

\maketitle

 {\bf Introduction.}
Anderson localization (AL), signifying the absence of wave spreading (diffusion) in a disordered medium \cite{anderson, Lee1985, Kramer1993, Evers2008, Lagendijk2009}, is a universal phenomenon in condensed matter physics. The Anderson transition (AT) occurs at a critical disorder strength, beyond which particle wave functions are confined within specific regions due to interference effects. Mobility edge (ME) is a pivotal concept in AL, representing the energy level at which the transition between localized and delocalized states takes place. Both the AT and the ME are indispensable for understanding wave behaviors in disordered materials, with broad implications, for instance, in metal-insulator transitions. Dimensionality plays a crucial role in the AT. As per the one-parameter scaling theory~\cite{DJThouless1974,Four,Hetenyi}, in one or two dimensions (2D), an infinitesimal weak disorder can lead to AL, while in 3D, the system becomes localized above a threshold of disorder strength.

Previous research on AL and ME has predominantly focused on Euclidean space or geometries.  Beyond these geometries, AL on infinite-dimensional lattice models such as the regular random graph (RRG) and its infinite-size limit, the Bethe lattice, has also been widely studied \cite{anderson73,zirnbauer,fyodorov,Altshuler1,Altshuler2,mirlin16,mirlin162,mirlin19prb,mirlin21,tarzia,tarzia2,parisi,sierant,Garel,prb2022}. Recently, hyperbolic lattices have emerged as a novel synthetic platform, with realizations in circuit quantum electrodynamics \cite{exp1, exp1b} and electrical-circuit networks \cite{exp2, exp3, edgestate3, edgestate4}. Formed by the regular tiling of the hyperbolic plane with constant negative curvature, hyperbolic lattices exhibit distinct crystalline symmetries \cite{crystal1, crystal2}. They host a plethora of exotic physical properties, such as high-genus Brillouin zone \cite{genus1, genus2, genus3}, uncharted topological phases \cite{edgestate1, edgestate2, edgestate3, edgestate4, edgestate5, edgestate6, edgestate7, edgestate8, edgestate9, edgestate10, Guohuaiming,Lux}, unusual flat bands \cite{flat1, flat2}, and set the stage for tabletop explorations of holography \cite{holo1, holo2, holo3, holo4, holo5, holo6, holo7}, curved-space quantum physics \cite{crystal2, curve2, curve3, curve4, curve5,curve6,curve7}, and efficient quantum error-correcting codes \cite{error1, error2, error3, error4, error5, error6}. By negating Euclid's parallel postulate, hyperbolic lattices feature a Hilbert space dimension exponentially growing with the layer (in contrast to the power-law dependence in Euclidean lattices) and harbor intricate closed loops pertinent to wave interference and particle transport. Therefore, when examining the influence of disorder on hyperbolic lattices, fundamental questions arise: Do AT and ME exist? How does the curvature of hyperbolic space influence AL?

In this work, we investigate the AL on hyperbolic lattices with random onsite disorders. Employing an inflation method, we generate hyperbolic lattices while preserving local connectivity at the outermost boundary to mitigate boundary effects. We take the $\{p,q\}=\{3,8\}$ and $\{p,q\}=\{4,8\}$ lattices as concrete examples and perform finite-size scaling of both the ratio of level spacings and inverse participation ratio (IPR) to determine the critical disorder strength of AT and the associated critical exponents. Furthermore, we relate the hyperbolic lattices to random regular graphs (RRGs) of the same connectivity, identifying the AT in the limit of $\{\infty, q\}$ via the cavity method. Our results indicate that AL occurs at strong disorder strengths on hyperbolic lattices, accompanied by the presence of MEs. And the transition points tend to increase with larger values of $\{p,q\}$ or curvature, ultimately saturating to that of the RRG at $p\rightarrow\infty$.

{\large{\bf Results}}

\begin{figure}[!t]
\centering
\includegraphics[width=3.33in]{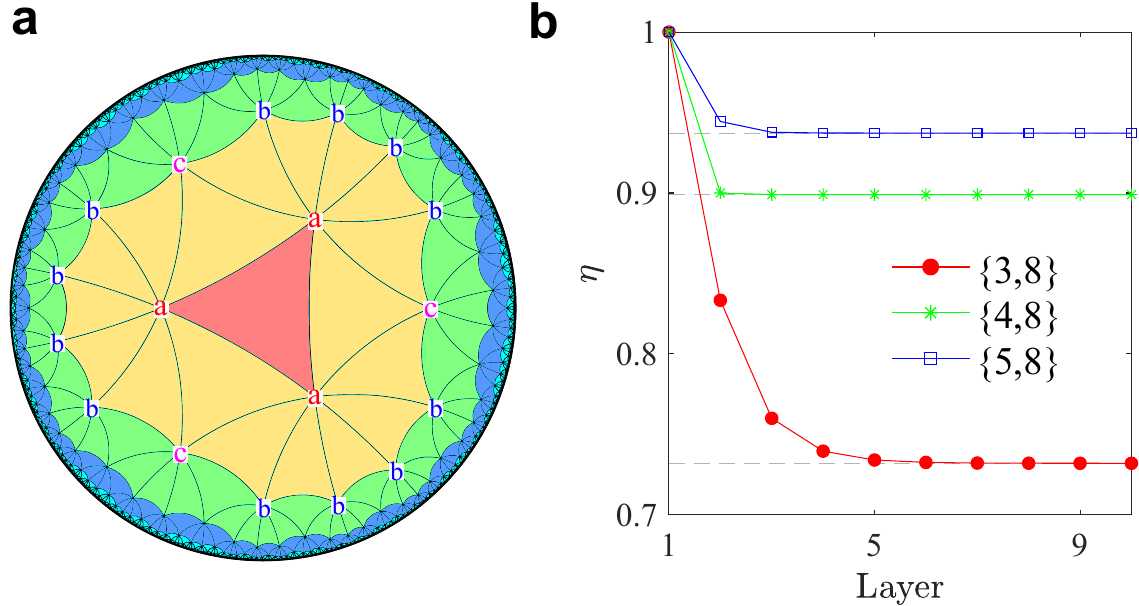}
\caption{{\bf Hyperbolic lattices visualized in the Poincaré disk.} {\bf{a}} Inflation pattern of the $\{3,8\}$ lattice, with each epoch (layer) marked in different colors. The vertices labeled by $a, b$ and $c$ denote sites with two, three, and four neighboring sites in each inflation epoch, respectively. {\bf{b}} The ratio of the number of outermost lattice sites to the total lattice sites as a function of the epoch, saturating to a finite value in the thermodynamic limit.}\label{fig1}
\end{figure}

{\bf{Hyperbolic lattices.}} In Schläfli symbol, a hyperbolic lattice $\{p,q\}$ refers to the tiling pattern of the hyperbolic plane with regular $p$-sided polygons meeting at a vertex, under the constraint $(p-2)(q-2)>4$. Here $q$ denotes the connectivity at each vertex. Such a tessellation is only achievable in the hyperbolic plane endowed with negative constant curvature given by \cite{edgestate1}:
\begin{eqnarray}
K=-\frac{p\pi}{A_{poly}}(1-\frac{2}{p}-\frac{2}{q}),
\end{eqnarray}
with $A_{poly}$ the area of the regular $p$-polygon. A larger value of either $p$ or $q$ corresponds to a larger (negative) curvature. A conventional way for visualizing the hyperbolic lattice is projecting it onto the Poincaré disk, as depicted in Fig. \ref{fig1}{\bf{a}}.

The generation of the hyperbolic lattice initiates from a central polygon, progressively adding neighboring polygons to the outermost edges in each epoch. Taking the $\{3, q\}$ lattice as an example, we start with a central triangle and proceed to tile the hyperbolic lattice layer by layer, with the latter one containing all the sites connected to the previous one. There exist three types of vertices, labeled as $a$, $b$, and $c$, respectively, corresponding to the lattice sites with two, three, and four neighbors in each epoch. The inflation pattern is governed by \cite{jahn} $a \mapsto b^{q-4}c, b\mapsto b^{q-5}c, c \mapsto b^{q-6}c$, yielding a recursive relation for the number of lattice sites in successive epochs:
\begin{align}
\begin{pmatrix}
N^b_n\\
N^c_n
\end{pmatrix}=
\begin{pmatrix}
q-5& 1\\
q-6& 1
\end{pmatrix}
\begin{pmatrix}
N^b_{n-1}\\
N^c_{n-1}
\end{pmatrix}.
\end{align}
Here $N^b_n (N^c_n)$ denotes the number of site $b$($c$) in the $n$-th layer. The larger eigenvalue of the above transfer matrix, i.e., $\lambda = \frac{1}{2}(\sqrt{q^2-8q+12}+q-4)$, determines the ratio of the number of sites at the outermost layer to the total lattice sites in the thermodynamic limit, $\eta_{n\rightarrow\infty} = (\lambda -1)/\lambda$. For the $\{3,8\}$ lattice, $\eta_{n\rightarrow\infty} = 0.732$. Similarly, we have $\eta_{n\rightarrow\infty} = 0.899$ for the $\{4,8\}$ lattice (see Supplementary Note 1). Fig. \ref{fig1}{\bf{b}} shows the variation of the ratio $\eta$ with the number of layers, and we observe that $\eta$ quickly converges to the value in the thermodynamical limit.

A finite boundary-to-bulk ratio indicates that the total number of lattice sites grows exponentially with the layer, which poses significant challenges for numerically approaching the thermodynamic limit. If we attempt to reach the thermodynamic limit using open boundary conditions (OBC), no AT occurs due to boundary effects (see Supplementary Note 2). To mitigate the boundary effects, our strategy is to randomly connect the sites at the outermost layer to preserve the local connectivity $q$. In the $p\rightarrow\infty$ limit, this treatment establishes an apparent connection to the RRG , which possesses AT and has a bulk-to-boundary ratio $\eta=1$. In contrast, two-dimensional Euclidean spaces with $\eta=0$ exhibit no localization transition. This suggests the AT point may increase as $\eta$ rises from 0 to 1. Intuitively, the hyperbolic lattices, which exhibit a local treelike structure with $0<\eta<1$, whose transition point may increase as $\eta$ increases, with the RRG as the limiting case.

\begin{figure}[!t]
	\centering
	\includegraphics[width=3.33in]{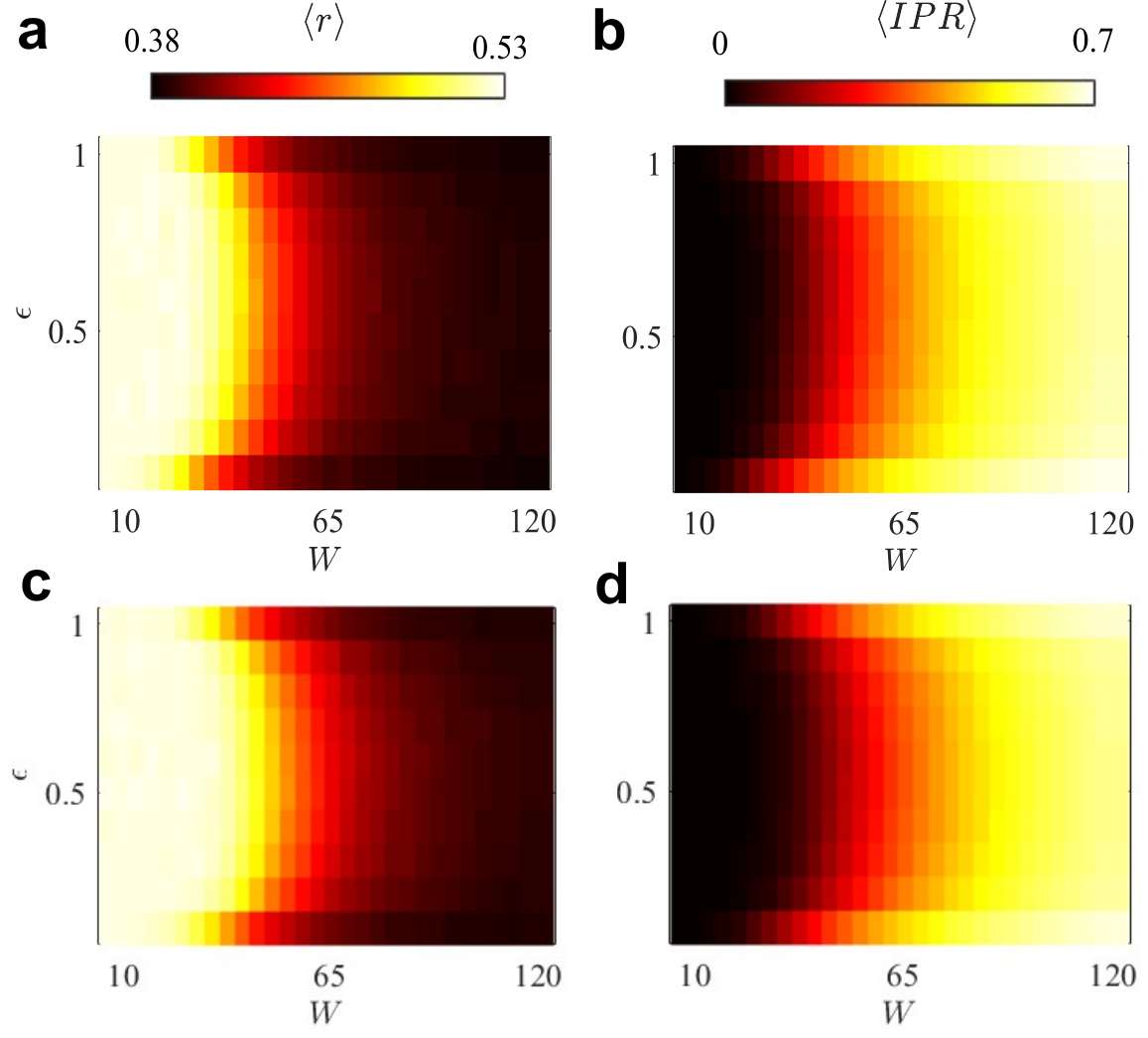}
	\caption{ {\bf The mobility edges.} The presence of mobility edges in the energy spectra of the hyperbolic $\{3,8\}$ lattice {\bf{a}},{\bf{b}} and $\{4,8\}$ lattice {\bf{c}},{\bf{d}}. {\bf{a}},{\bf{c}} The average ratio of level spacings $\langle r\rangle$ and {\bf{b}},{\bf{d}} average inverse participation ratio $\langle IPR\rangle$ with respect to the disorder strength and eigenenergies.  The eigenenergies are divided into ten equal segments and rescaled to the interval $\epsilon\in[0,1]$. The averages are taken within each segment and over $1000$ disorder realizations. }\label{fig2}
\end{figure}
		{\bf{ Emergence of mobility edge.} }  We now examine the impact of disorder on the hyperbolic lattice using the Hamiltonian
\begin{align}
H=\sum_{<i,j>}t(c_i^\dagger c_{j} +h.c.)+\sum_j\epsilon_j c_j^\dagger c_j.
\end{align}
Here $c_j^\dagger (c_j)$ represents the creation (annihilation) operator on site $j$. $t$ is the hopping strength between neighboring sites and set to the energy unit $t=1$. The onsite disorder $\epsilon_j$ is uniformly distributed in the interval $[-W/2, W/2]$  (see Supplementary Note 3). We employ two quantities to characterize the AT on hyperbolic lattices. The first one is the ratio of level spacings, defined as \cite{Shkloyskii1993, Oganesyan2007}:
\begin{align}
r_i = \frac{\min(\delta_i, \delta_{i-1})}{\max(\delta_i, \delta_{i-1})},
\end{align}
where $\delta_i = E_i - E_{i-1}$ with energy levels ${E_i}$ arranged in ascending order. In the localized (delocalized) phase, the spectrum statistic satisfies a Poisson distribution with $\langle r\rangle \approx 0.387$ (Wigner-Dyson distribution with $\langle r\rangle \approx 0.529$) \cite{Shkloyskii1993, Oganesyan2007, Atas2013,mondaini, xiaopeng}. Here $\langle r\rangle$ is the averaged ratio over energy levels and disorder realizations. The second quantity is the IPR \cite{DJThouless1974, Bell1972, Wegner,Schreiber,Hashimoto1992}. Given an eigenstate $|\psi_m\rangle$, it is defined as $IPR = \sum_j|\psi_{m,j}|^4$, with $\psi_{m,j}$ being the spatial component. For extended states and localized states, the IPR tends to $0$ and a nonzero finite value, respectively.

\begin{figure}[!t]
	\centering
	\includegraphics[width=3.33in]{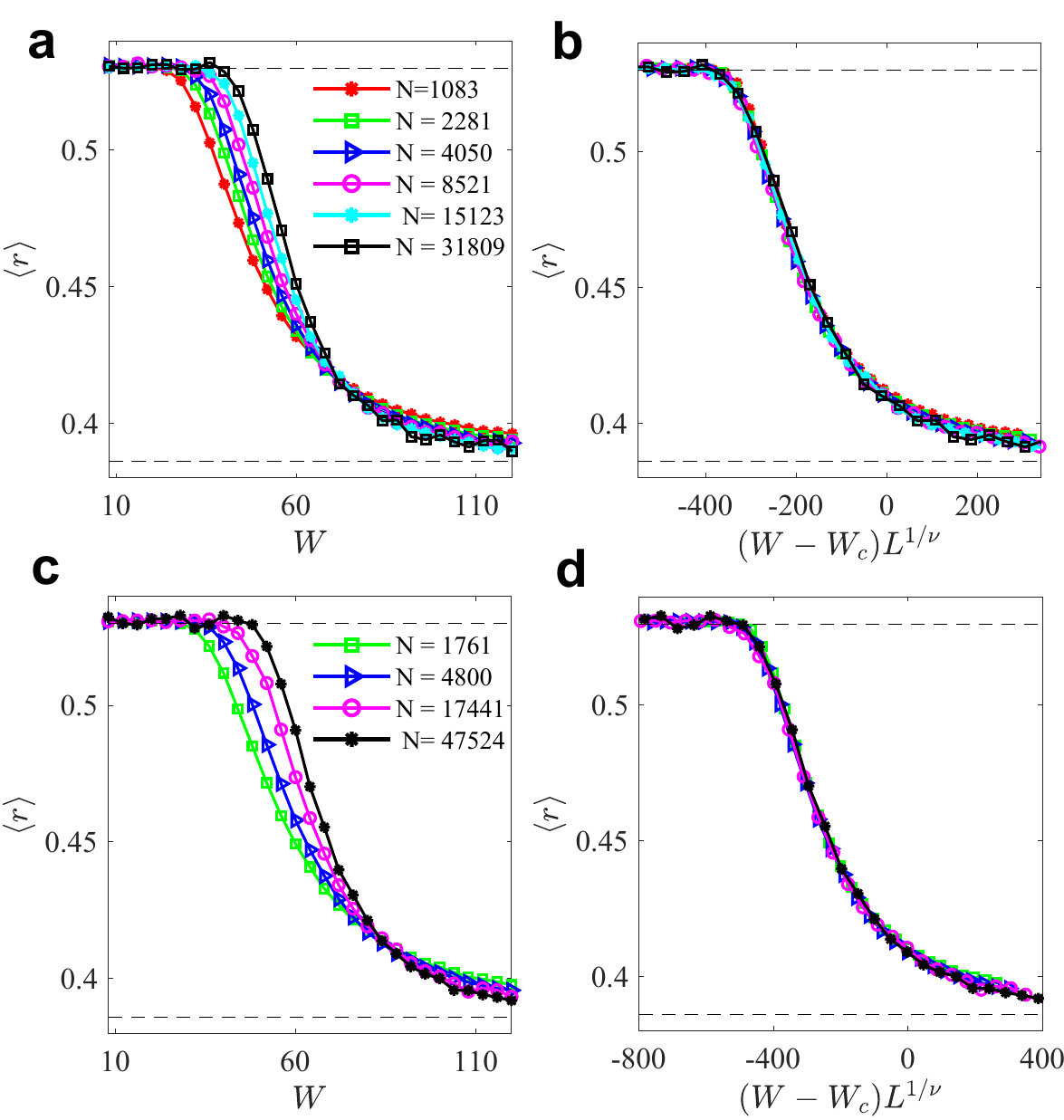}
	\caption{{\bf Finite-size scaling of the level statistics.}The average ratio of level spacings $\langle r\rangle$ as a function of the disorder strength $W$ with different system sizes for {\bf{a}} $\{p, q\} = \{3,8\}$ and {\bf{c}} $\{p, q\} = \{4,8\}$ lattice. {\bf{b}},{\bf{d}}, Finite-size scaling of {\bf{a}},{\bf{c}}, $L={\rm ln}N$ and $N$ is the number of sites. In {\bf{b}}, the transition point $W_c=77.2\pm3$ and the critical exponent $\nu=1.02\pm0.05$. In {\bf{d}}, $W_c=88.2\pm3$ and the critical exponent $\nu=0.95\pm0.05$. The average is taken over the central $1/10$ eigenstates with $20\sim 1000$ disorder realizations. The horizontal lines mark the value $\langle r\rangle=0.387$ and  $\langle r\rangle=0.529$.}\label{fig3}
\end{figure}

To discern the appearance of MEs with increasing disorder strength, we partition the energy spectrum into ten segments (each containing an equal number of states) and consider the characteristics of eigenstates in different energy windows.  We rescale the energy spectrum as $\epsilon = C(E)/N$, where $C(E)$ represents the number of eigenstates below energy $E$, and $C(E_{\text{max}}) = N$, with $E_{\text{max}}$ the energy of the highest excited state.  Averaging over both the eigenstates within each segment and the disorder samples, we present the results of $\langle r\rangle$ and $\langle IPR\rangle$ for the hyperbolic $\{3,8\}$ lattice ($\{4,8\}$ lattice) in Figs.\ref{fig2}{\bf{a}},{\bf{b}} ( Figs. \ref{fig2}{\bf{c}},{\bf{d}}), respectively. The extended-localized transition points differ for the central and side regions of the spectrum, indicating the presence of MEs.
 A more sophisticated finite-size analysis of $\{3,8\}$ lattice suggests that the AT occurs at around $W_c = 54$ at the band bottom  (see Supplementary Note 4 and 5) and around $W_c = 77.2$ at the band center [See Fig. \ref{fig3}]. Our numeric verifies the existence of MEs for other $\{p,q\}$ cases. For a fixed disorder strength $W$ which slightly smaller than $W_c$, the mobility edge divides the energy spectrum into three regions: two side regions with localized states and a central region with extended states. As $p$ increases (curvature increases), the mobility edge shifts towards the spectrum boundaries, resulting in the shrinking of the localized regions at the spectrum edges.

{\bf{Finite-size scaling of the level statistics and IPR.}} In the subsequent discussions, we focus on the eigenstates at the band center (middle $1/10$). When the AT occurs there, all eigenstates should become localized due to the presence of MEs. To precisely determine the transition point and extract the associated critical exponents, we perform the finite-size analysis below.

First for the level statistics. In Fig. \ref{fig3}{\bf{a}}, we present the variation of $\langle r\rangle$ with respect to the disorder strength for different system sizes of the $\{3,8\}$ lattices, ranging up to $N=31809$. The value decreases from $\langle r\rangle=0.529$ (delocalized region) at small disorder to $\langle r\rangle=0.387$ (localized region) at large disorder. Noting that the crossing points of neighboring $\langle r\rangle$-curves shift toward larger disorder strength for larger system sizes. To pinpoint the AT, we employ the fitting $\langle r\rangle=f[(W-W_c)L^{1/\nu}]$. Here $L={\rm ln}N$, $W_c$ denotes the transition point and $\nu$ represents the critical exponent. Our numeric indicates that the best fit are $W_c =77.2\pm3$ and $\nu = 1.02\pm0.05$, as depicted in Fig. \ref{fig3}{\bf{b}}. Similarly for the $\{4,8\}$ lattices with system size up to $N=47524$, the finite-size scaling yields $W_c = 88.2\pm3$, $\nu = 0.95\pm0.05$, as illustrated in Figs. \ref{fig3}{\bf{c}},{\bf{d}}.
\begin{figure}[!t]
	\centering
	\includegraphics[width=3.33in]{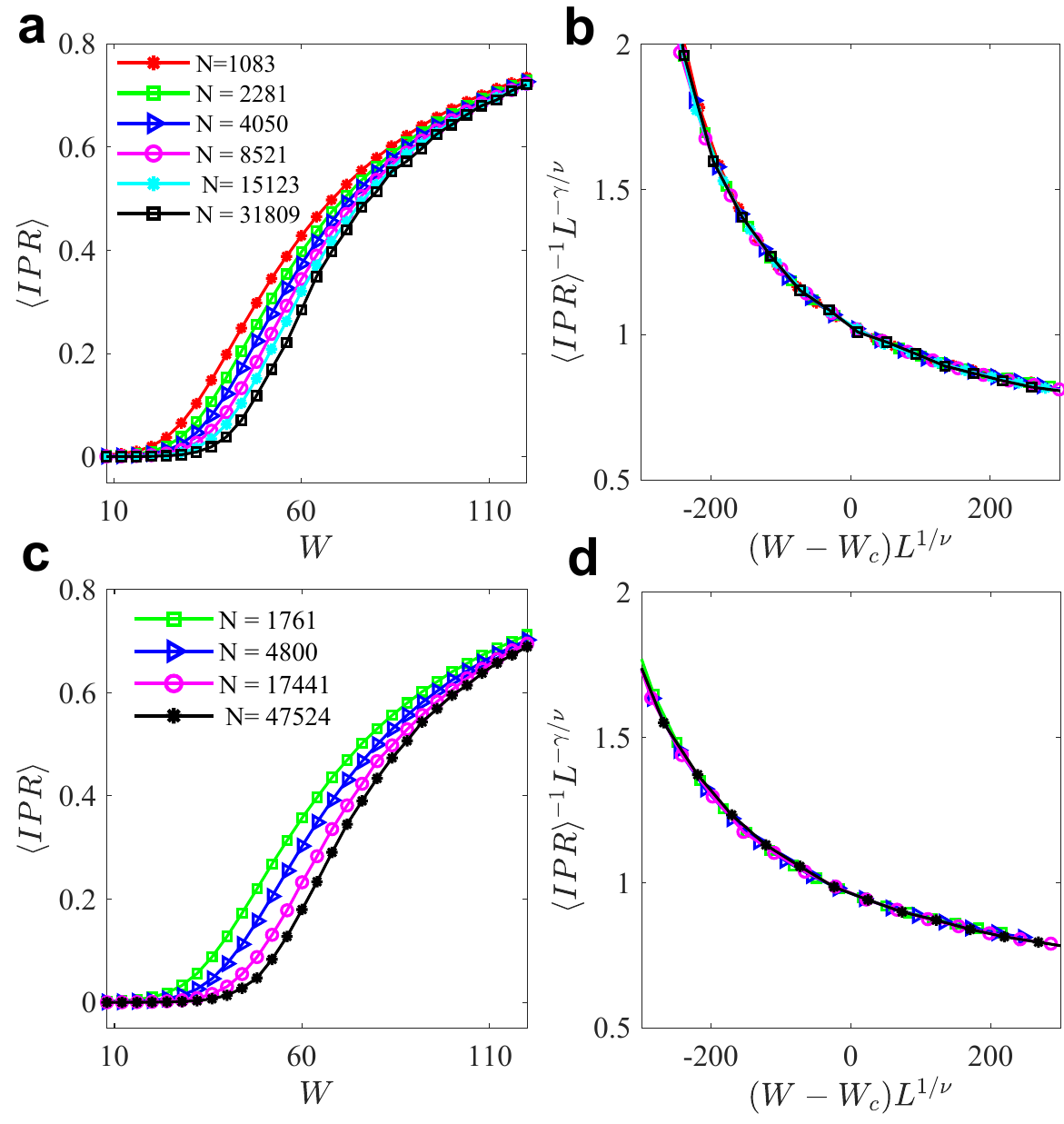}
\caption{{\bf Finite-size scaling of inverse participation ratio.} The average inverse participation ratio $\langle IPR\rangle$ as a function of the disorder strength $W$ with different system sizes for {\bf{a}} $\{p, q\} = \{3,8\}$ and {\bf{c}} $\{p, q\} = \{4,8\}$ lattice. {\bf{b}},{\bf{d}} Finite-size scaling of {\bf{a}},{\bf{c}}, $L={\rm ln}N$. In {\bf{b}}, the transition point $W_c =83.0\pm 4$ and the critical exponent $\gamma = 0.25\pm0.03$. In {\bf{d}}, $W_c =94\pm 4$ and $\gamma = 0.25\pm 0.03$. The average is taken over the central $1/10$ eigenstates with $20\sim 1000$ disorder realizations.}\label{fig4}
\end{figure}

Second for the IPR. Across the AT, we observe a change of the IPR from $0$ to a finite value with increasing disorder strength, as depicted in Figs. \ref{fig4}{\bf{a}},{\bf{c}} for the $\{3,8\}$ and $\{4,8\}$ lattices, respectively. For the IPR, we employ the fitting function $\langle IPR\rangle^{-1}L^{-\gamma/\nu}=g[(W-W_c)L^{1/\nu}]$~\cite{Hashimoto1992,YWang2016}, where $\nu$ is the same critical exponent obtained in the analysis of $\langle r\rangle$, and $\gamma$ is another critical exponent. Our numeric suggests the optimal fit of IPR corresponds to $W_c =83.0\pm4$, and $\gamma = 0.25\pm 0.03$ for the $\{3,8\}$ lattice, and $W_c =94\pm4$, and $\gamma = 0.25\pm 0.03$ for the $\{4,8\}$ lattice. As illustrated in Figs. \ref{fig4}{\bf{b}},{\bf{d}}, we observe a perfect overlap the IPR curves with different system sizes under this scaling. In addition,
 the eigenstates of disordered systems near the transition point exhibit fractal behavior, $\langle IPR \rangle\sim L^{-D_2}$, which characterized by fractal dimensions $D_2$.  The fractal dimensions versus  of the  $\{3,8\}$ ($\{4,8\}$) lattice around the transition point are $D_2^{\{3,8\}} = 0.3\pm 0.02$  and  $D_2^{\{4,8\}} = 0.28\pm0.03$  ( see Supplementary Note 6). 

For both the $\{3,8\}$ and $\{4,8\}$ lattices, the transition points extracted from the finite-size scaling of $\langle r\rangle$ and IPR align with each other. It is worth noting that the critical exponent $\nu\approx 1$ for both lattices, differing from the 3D AT induced by random disorder, which has a known critical exponent of $\nu \approx 1.58$ \cite{Mackinnon1994,Slevin1999,Slevin2014}. Instead, it is rather close to the AT in 1D and 2D induced by quasiperiodic potentials, whose critical exponent is also $\nu\approx 1$ \cite{Hashimoto1992,YWang2016,AA,YuWang2023}. This suggests that there may be similar critical behaviors near the transition points for both the hyperbolic lattices and quasiperiodic lattices. It is worthy to note that an alternative two-sided scaling analysis \cite{AT2,prb2022} can be performed to yield rather close critical disorder strengths for both cases  (see Supplementary Note 7).  

\begin{figure}[!t]
\centering
\includegraphics[width=3.33in]{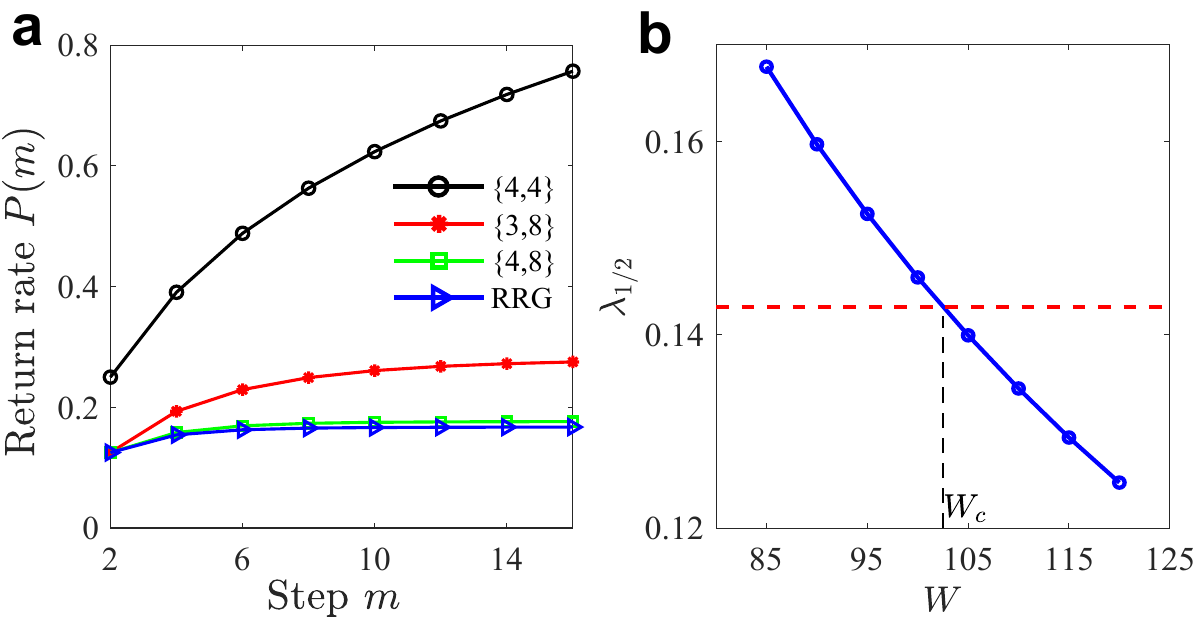}
\caption{{\bf Localization phase transition on random regular graph (RRG) with connectivity $D = 8$.} {\bf{a}} The expected return rate $P(m)$ of a random walker within $m$ steps as a function of $m$ on hyperbolic $\{3,8\}$ (red), $\{4,8\}$ (green), RRG (blue) of $D=8$, and Euclidean $\{4,4\}$ lattice, respectively. The hyperbolic $\{p,8\}$ case converges to a finite value of expected return for large $m$ and approaches that on the RRG as $p\rightarrow\infty$. {\bf{b}} The eigenvalue $\lambda_{1/2}$ of the integral operator defined in Eq. (\ref{eigen}) (blue line) as a function of the disorder strength $W$. Its intersection with $\lambda_{1/2} = 1/7$ (dashed red line) around $W_c = 102.5$ marks the transition point on the RRG in the thermodynamic limit.}\label{fig5}
\end{figure}
{\bf{Anderson transition in the limiting case.}}   The finite-size scaling analysis implies that the $\{4,8\}$ lattice with larger curvature exhibits a larger critical disorder strength of AT compared to the $\{3,8\}$ case. To seek a qualitative understanding of the variation of the AT on $\{p,q\}$ lattices with fixed connectivity of $q$, we note that particle transport and AL are intimately linked to the wave interference induced by closed loops on the lattice \cite{loop}. As $p$ increases, both the curvature and loop size grow, while the local treelike structure of the lattice remains intact. When $p\rightarrow\infty$, the loop size diverges, akin to the RRG. Consequently, the transition point with larger $p$ should saturate to that of the RRG with connectivity $D=q$. A pertinent analysis involves the expectation value $P(m)$ of a particle undergoing random walks on the lattice and returning to the original position within $m$ steps, as illustrated in Fig. \ref{fig5}{\bf{a}}. With increasing $m$, it becomes evident that $P(m)$ on hyperbolic lattices converges to a finite value \cite{AT2}, and the $P(m)$ curve saturates to that of the RRG as $p\rightarrow\infty$. In contrast, for a normal 2D Euclidean lattice, e.g., $\{4,4\}$, this value eventually diverges. This observation aligns with the fact that AT occurs at infinitesimal weak disorder in 2D Euclidean lattices, whereas it occurs at a pronounced disorder strength on hyperbolic lattices and the random regular graph. 

While the AT on the RRG can be assessed from the finite-size scaling, here, we determine the transition point in the thermodynamic limit using the cavity method. 
 Recently, G. Parisi, et.al.  proposed a new method to approch the transition point on Bethe lattices\cite{parisi}. Instead of obtaining the transition point through the study of the imaginary part of the self-energy \cite{ prb2022,sierant,anderson73,mirlin19prb}, they introduce a novel criterion for the localization transition based on the stability of the population of the propagators. We identify the transition point of RRG with a connectivity of $q=8$ by following this method. 

 Formally, for the RRG with connectivity $D=q$, the diagonal element of the resolvent $\mathcal{G}(E)=1/(E-H)$ is given by $\mathcal{G}_{ii}(E)=1/({E-\epsilon_i - \sum_{k=1}^{D}G_k(E)})$ \cite{anderson, sierant, parisi,prb2022,anderson73,mirlin19prb}, where the summation is over the $D$ neighbors of site $i$. $G_{k}$ represents the cavity Green’s function and is treated as independent, identically distributed random variable. It corresponds to the Hamiltonian restricted to a sub-tree with the site $k$ removed. The cavity Green’s functions satisfy the self-consistent condition:
\begin{align}
G_l(E)=\frac{1}{E-\epsilon_l - \sum_{k=1}^{D-1}G_k(E)}.\label{eq:self}
\end{align}
The probability distribution $P(G)$ can be evaluated via the pool method (See Ref. \cite{Garel} and the Methods section). Subsequently, we take the kernel function \cite{parisi,sierant} $K_s(x,y) = \int_{-W/2}^{W/2}{\rm{d}}\epsilon\frac{|x|^{2s}}{Wy^2} \mathcal{P}_\zeta(\frac{1}{y}-\epsilon-x)$, where $\mathcal{P}_\zeta(x)$ represents the distribution of $\zeta = \sum_{j=1}^{D-2}G_j$. The objective is to find the largest eigenvalue of its integral operator\cite{parisi},
\begin{align}
\int {\rm{d}}x K_s(y,x) \phi_s(x)=\lambda_s\phi_s(y). \label{eigen}
\end{align}
The AT condition is given by $(D-1)\lambda_{s=1/2} =1$. Fig. \ref{fig5}{\bf{b}} plots the dependence of the eigenvalue $\lambda_{1/2}$ on the disorder strength $W$. For the $D=q=8$ case, the critical disorder strength thus obtained is $W_c=102.5$, in agreement with the numerical result $W_c = 105$ reported in Ref. \cite{tarzia2}.

{\large{\bf{Discussion}}}

 In conclusion, we delve into the AT on hyperbolic lattices with random onsite disorders, unveiling the presence of MEs. Taking the $\{3,8\}$ and $\{4,8\}$ lattices as examples, we pinpoint the critical disorder strengths and critical exponents through finite-size scaling analysis of both the average ratio of level spacings and IPR. Furthermore, we draw the connection between the $p\rightarrow\infty$ case and the RRG, determining the AT in the limiting scenario through the cavity method. Our findings indicate that the transition point escalates with increasing $\{p,q\}$ or curvature, ultimately saturating towards that of the RRG of the same connectivity.

Our results indicate that AL on hyperbolic lattices occur at large disorder strengths. This is in stark contrast to the 2D Euclidean lattices, where even very weak disorder can induce AL in the absence of spin-orbit coupling or time-reversal symmetry breaking according to the scaling theory. This underscores the need for a new scaling theory capable of accurately describing AL in non-Euclidean spaces. Additionally, our results unveil noteworthy similarities in the critical exponents between the AL on hyperbolic lattices and quasiperiodic systems, suggesting possible commonalities between the two. Our results raise many interesting issues that warrant further in-depth study. Specifically, whether the AL and MEs are stable in the presence of interactions, namely, are there many-body localization and many-body ME phenomena in hyperbolic lattices? Given that disorder is pervasive, our findings hold relevance in the recent implementations of hyperbolic lattices in circuit quantum electrodynamics \cite{exp1, exp1b} and electrical-circuit networks \cite{exp2, exp3, edgestate3, edgestate4}. Our findings set the stage for future investigations into the localization phenomena in hyperbolic lattices.\\

 We note that during the finalization of this paper, two independent preprints \cite{curve6,AT2} exploring AL on disordered hyperbolic lattices appeared online. 
	In Ref.  \cite{curve6}, it was shown that negative curvature manifolds feature a cutoff for the probability of self-returning paths, diminishing the significance of interference effects and reinstating classical diffusive behavior through explicit calculations of the Cooperon, which is directly linked to weak-localization corrections. These results imply that localization on negative curvature manifests at a finite disorder strength. 
	Instead of investigating the Cooperon on negative curvature manifolds, we (also Ref. \cite{AT2}) examine discrete lattice models, specifically disordered hyperbolic lattice models. Our numerical findings indeed indicate that the localization transition occurs at strong disorder strengths. 
	Although some of our findings bear resemblance to those reported in Ref. \cite{AT2}, our work has significant differences.
	To mitigate boundary effects, we used different boundary conditions. While Ref. \cite{AT2} adopts periodic boundary conditions (PBC), we have opted for a randomized connection of boundary sites. Implementing PBC requires complex crystallographic analysis of bond connections, which also varies for different hyperbolic lattices. In contrast, our random connection protocol is more experimentally friendly. And on the other hand, the relationship to random regular graphs as $p\rightarrow \infty$ is more apparent in our framework.
We have employed a different method to analyze the transition points and critical exponents. The critical exponents derived from these approaches differ and carry distinct physical interpretations.
	 Additionally, our paper investigates mobility edges on hyperbolic lattices, a topic not addressed in the reference. Furthermore, we have identified the Anderson transition for RRG (the limiting scenario as $p\rightarrow \infty$) utilizing the cavity method.\\

{\large \bf Methods}

For an RRG with $N$ lattice sites, the typical loop size is approximately ${\rm ln}N$, which diverges in the thermodynamic limit. In this scenario, any portion of the RRG can be regarded as a Bethe lattice with a local treelike structure. Utilizing the cavity method, we can deduce the self-consistent equations of Green’s functions. Then we introduce the pool method \cite{Garel} to extract their probability distributions. Further more, based on the stability of the population of propagators, and introduce the kernel function,  a criterion for the localization–delocalization transition can be provided\cite{parisi}.

{\bf{Cavity method.}}  Let us consider the tight-binding Hamiltonian defined on an RRG\cite{greenbook}:
 \begin{align}
H =& H_0+H_1, \\\nonumber
H_0 =& \sum_n \epsilon_n|n\rangle\langle n|,\\\nonumber
H_1 =& \sum _{\langle m,n\rangle} V|m\rangle\langle n|, \label{H}
\end{align}
where the symbol $\langle m,n\rangle$ denotes the site $m$ and $n$ are interconnected on the graph. Define the Green's function $\mathcal{G}(z) \equiv \frac{1}{z-H}= \frac{1}{z-H_0-H_1}$, and $G_0(z) \equiv \frac{1}{z-H_0}$. We have $\mathcal{G} = G_0+G_0 H_1 G_0+G_0H_1G_0H_1G_0+...$, and their matrix elements satisfy
\begin{align}
\mathcal{G}_{mn}&\equiv G(m,n)=\\\nonumber
& G_0(m,n)+\sum_{l_1,l_2}G_0(m,l_1)\langle l_1|H_1|l_2\rangle G_0(l_2,n) +\\\nonumber
&\sum_{l_1,l_2,l_3,l_4}G_0(m,l_1)\langle l_1|H_1|l_2\rangle G_0(l_2,l_3)\langle l_3|H_1|l_4\rangle G_0(l_4,n)\\\nonumber
&+....
\end{align}
Notice that $G_0(l_1,l_2) = \delta_{l_1,l_2}G_0(l_1)$, $G_0(l_1)\equiv (z-\epsilon_{l_1})^{-1}$, and $\langle l_1|H_1|l_2\rangle=V\delta_{\langle m,n\rangle}$, where $\delta_{l_1,l_2}$ is the Kronecker's delta function, and $\delta_{\langle m,n\rangle} = 1~(0)$ if the site $m$ and $n$ are interconnected (not connected). Thus we have
\begin{align}
&G(m,n)=\delta_{m,n}G_0(m)+G_0(m) V G_0(n)\delta_{\langle m,n\rangle}\\\nonumber
&+\sum_{l_1}G_0(m)VG_0(l_1)V G_0(n)\delta_{\langle m,l_1\rangle}\delta_{\langle l_1,n\rangle}\\\nonumber
&+\sum_{l_1,l_2}G_0(m)VG_0(l_1)VG_0(l_2)V G_0(n)\delta_{\langle m,l_1\rangle}\delta_{\langle l_1,l_2\rangle}\delta_{\langle l_2,n\rangle}\\\nonumber
&+....
\end{align}
Each term in the above equation can be represented by a diagram, starting from the site $m$ and ending at the site $n$. Recombining of all the diagrams, the above equation can be rewritten as
\begin{align}
G(m,n)=&\sum G(m,m) V G(l_1,l_1[m]) V G(l_2,l_2[m,l_1])V\\\nonumber
&...VG(n,n[m,l_1,l_2,...]),
\end{align}
where the summation is taken over all self-avoiding paths starting from the site $m$ and ending at the site $n$. $G(m,n[l,j,..])$ is the element of the Green's function of the new Hamiltonian originate from Eq. (\ref{H}) but with sites $l,j,...$ removed. In particular, the diagonal element $G(m,m)$ is
\begin{align}
&G(m,m)=G_0(m)+\\\nonumber
&\quad\sum G(m,m) V G(l_1,l_1[m]) V G(l_2,l_2[m,l_1])V...VG_0(m)\\\nonumber
&\quad\quad\quad\quad=G_0(m)+ G(m,m) \Delta(m) G_0(m)
\end{align}
with
\begin{align}
\Delta(m)\equiv \sum V G(l_1,l_1[m]) V G(l_2,l_2[m,l_1])V...V
\end{align}
the self-energy. Then we have
\begin{align}
G(m,m;z)=\frac{G_0(m)}{1-G_0(m)\Delta(m)}=\frac{1}{z-\epsilon_m - \Delta(m)}.
\end{align}
For the Bethe lattice with connectivity $D$, the self-energy takes the simple form (due to the lack of loops) 
\begin{align}
\Delta(m) = V^2\sum_{n=1}^{D} G(n,n[m]).
\end{align}
Without loss of generality, we set $V =1$, then the diagonal element of the resolvent $\mathcal{G}(E)=\frac{1}{E-H}$ is 
\begin{align}
\mathcal{G}_{ii}(E)=\frac{1}{E-\epsilon_i - \sum_{k=1}^{D}G_k(E)}.\label{self1}
\end{align}
The summation is over the $D$ neighbors of the site $i$, and  $G_{k}\equiv G(k,k[i])$ is the cavity Green's function. In the thermodynamic limit, the set of Green's functions $G_{k}$ ($k=1,2,...,D$) can be regarded as random numbers following the same distribution $P(G)$. It can be checked that they satisfy the self-consistent equation:
\begin{align}
G_l(E)=\frac{1}{E-\epsilon_l - \sum_{k=1}^{D-1}G_k(E)}.\label{self2}
\end{align}
It is noteworthy that despite being derived from the Bethe lattice, Eqs. (\ref{self1}) and (\ref{self2}) remain applicable in the thermodynamic limit for the RRG because short loops on RRG barely exist, as elaborated in Ref. \cite{parisi}.

{\bf{Pool method.}} One can obtain the distribution $P(G)$ through the pool method \cite{Garel}. The procedure is listed below (we focus on the scenario $E =0$):
\quad\\
(1) Draw randomly from uniform distribution $[-1,1]$ and generate the initial pool $\{G^{(1)}_0, G^{(2)}_0,...,G^{(M_{pool})}_0\}.$\\
(2) Randomly draw $D-1$ numbers from the initial pool and $\epsilon\in(-W/2,W/2)$; Calculate $G_1^{(i)}=\frac{1}{-\epsilon-\sum_{i_k=1}^{D-1}G^{(i_k)}_0}$.\\
(3) Repeat step-2 $M_{pool}$ times and obtain the first-generation pool $\{G^{(1)}_1, G^{(2)}_1,...,G^{(M_{pool})}_1\}.$\\
(4) Randomly draw $D-1$ numbers from the $(n-1)$th pool and $\epsilon\in(-W/2,W/2)$; Calculate $G_n^{(i)}=\frac{1}{-\epsilon - \sum_{i_k=1}^{D-1}G^{(i_k)}_{n-1}}$.\\
(5) Repeat step-4 $M_{pool}$ times and obtain the $n$-th-generation pool $\{G^{(1)}_n, G^{(2)}_n,...,G^{(M_{pool})}_n\}.$\\
(6) Repeat step-4 and step-5 $N-1$ times and obtain the $N$-th offspring $\{G^{(1)}_N, G^{(2)}_N,...,G^{(M_{pool})}_N\}.$

The statistics of the $N$-th pool should converge to the distribution $P(G)$ as $N$ and $M_{\text{pool}}$ approach infinity. 

{\bf{The kernel function.}}  Once armed with the probability distribution $P(G)$, a more nuanced analysis is required to pinpoint the localization phase transition point of the RRG. This involves considering the shortest individual path $p$ from $i$ to $j$ with a length of $L$ and defining the susceptibilities, as detailed in \cite{parisi}:
\begin{align}
\chi_p = \frac{\partial G_j}{\partial \epsilon_i}=(\Pi_{l=1}^L\frac{\partial G_l+1}{\partial G_l})\frac{\partial G_i}{\partial \epsilon_i}=\Pi_{l=1}^LG_l^2.
\end{align}
The $s$th moment of $\chi$ (For brevity, the index $p$ is omitted) is
 \begin{align}
\langle\chi^s\rangle = \int_0^{\infty}{\rm d}\chi Q(\chi)\chi^s = \langle \Pi_{l=1}^L G_l^{2s}\rangle \equiv C_L \lambda_s^L.
\end{align}
The localization phase transition is determined by the condition \cite{parisi}:
\begin{align}
(D-1)\lambda_{s=1/2} =1.\label{eq:cond}
\end{align}

It can be shown that $\lambda_s$ is related to the largest eigenvalue of an integral kernel function. We focus the physics around $E = 0$ and note that the propagators along the path $p$ fulfill 
\begin{align}
G_{l+1}=-\frac{1}{\epsilon_{l+1}+G_l+\zeta},
\end{align}
where $\zeta=\sum_{j=1}^{D-2}G_j$, and $G_j$ are i.i.d. variables with distribution $P(G)$. Thus we obtain the conditional probability
\begin{align}
&P(G_{l+1}|G_l) =\\\nonumber
 &\int_{-W/2}^{W/2}{\rm{d}}\epsilon\int_{-\infty}^{\infty}{\rm{d}}\zeta \frac{1}{W} \mathcal{P}_\zeta(\zeta)\delta(G_{l+1}+\frac{1}{\epsilon_{l+1}+G_l+\zeta}).
\end{align}
Define the kernel $K(y,x) \equiv P(y|x)$, we have
\begin{align}
\langle \chi^s\rangle =& \langle G_L^{2s}...G_1^{2s}\rangle\\\nonumber
=& \int {\rm{d}}G_{L}...{\rm{d}}G_0 G_L^{2s}K(G_{L},G_{L-1})...G^{2s}_1K(G_1,G_0) P(G_0).
\end{align}
We can view the above integral from the perspective of transfer matricx by discretizing the integrals, and define the kernel function by
\begin{align}
K_s(y,x) \equiv K(y,x)x^{2s},
\end{align}
which takes the form of
\begin{align}
K_s(x,y) = \int_{-W/2}^{W/2}{\rm{d}}\epsilon\frac{|x|^{2s}}{Wy^2} \mathcal{P}_\zeta(\frac{1}{y}-\epsilon-x).
\end{align}
Here we have used the fact that $K(x,y) = \frac{1}{y^2}K(\frac{1}{y},x)$ from the properties of Dirac $\delta$ function. $\mathcal{P}_\zeta(\zeta)$ denotes the distribution of $\zeta = \sum_{j=1}^{K-1}G_j$. Finally, we determine $\lambda_s$ by identifying the largest eigenvalue of its integral operator,
\begin{align}
\int {\rm{d}}x K_s(y,x) \phi_s(x)=\lambda_s\phi_s(y).\label{eigenS}
\end{align}

To numerically evaluate the eigenvalue of Eq. (\ref{eigenS}), we introduce a cutoff $x_M$ for the integral and discretize the interval $(x_1, x_2, ..., x_M)$. We then introduce the discrete basis $\frac{1}{\Delta_i^{1/2}}\delta_{x_i}(x)$, where $\delta_{x_i}(x) = 1$ for $x \in [x_i, x_i + \Delta_i]$ (and 0 otherwise), with $\Delta_i = x_{i+1} - x_i$. In this basis, $\phi_s(x)$ can be vectorized as $\phi_s(x) = \sum_{i=1}^{M}c_i \frac{1}{\Delta_i^{1/2}}\delta_{x_i}(x)$, where $c_i = \frac{1}{\Delta_i^{1/2}}\int_{-\infty}^{\infty}{\rm{d}}x \delta_{x_i}(x) \phi_s(x)$ represents the coefficients. Under this basis, the integral operator can be approximated by an $M\times M$ matrix. Finally, we perform exact diagonalization on this matrix to obtain the largest eigenvalue $\lambda_s$. For the RRG with connectivity $D =  8$, the localization transition condition is expressed as $\lambda_{1/2}(W_c) = 1/7$, with $W_c$ the critical disorder strength in the thermodynamic limit. As illustrated by in Fig. 5{\bf{b}} of the main text, the transition point thus determine is $W = 102.5$, which closely matches the value ($W_c = 105$) reported  in Ref. \cite{tarzia2}. \\

{\large{\bf Data availability}} \\
The data that support the findings of this study are available at https://doi.org/10.5281/zenodo.13899656.\\

{\large{\bf Code availability}} \\
The code that support the findings of this study are available at https://doi.org/10.5281/zenodo.13899656.\\

{\large{\bf References}}

\clearpage


\renewcommand{\figurename}{Supplementary Figure}
\setcounter{figure}{0}
\setcounter{table}{0}
\pagebreak
\widetext
\begin{center}
\textbf{\large  Supplemental Material for ``Anderson transition and mobility edges on hyperbolic lattices with randomly connected boundaries''}
\end{center}

In this Supplemental material, we provide more details on (I) the inflation method of generating the hyperbolic $\{4,8\}$ lattice;  (II) Level statistics under open boundary conditions; (III) the spectrum of hyperbolic Hamiltonian;  (IV) Finite-Size scaling analysis methods; (V) Finite-size scaling analysis of Anderson localization (AT) at band bottom; (VI) Fractal dimensions and (VII) Two-sided scaling analysis.

\section{Supplementary Note 1: \uppercase{ Inflation of the hyperbolic $\{4,8\}$ lattice}}
Similar to the inflation procedure of the \{3,q\} lattice in the main text, we generate the \{4,q\} hyperbolic lattice step by step as sketched in Supplementary Figure \ref{48}. The inflation pattern is given by: 
\begin{align}
a \mapsto a^{q-3} b^{q-2};\quad  b\mapsto a^{q-4} b^{q-3}.
\end{align}
Here $a, b$ represent lattice sites possessing two and three neighboring sites in each layer or epoch, respectively. Then we have the following recursive relation ($n>1$):
\begin{align}
\begin{pmatrix}
N^a_n\\
N^b_n
\end{pmatrix}=
\begin{pmatrix}
q-3& q-2\\
q-4& q-3
\end{pmatrix}
\begin{pmatrix}
N^a_{n-1}\\
N^b_{n-1}
\end{pmatrix},
\end{align}
with $N^a_n (N^b_n)$ the number of $a$ ($b$) type sites in the $n$-th layer. The largest eigenvalue of the above transfer matrix is 
\begin{align}
\lambda = \sqrt{q^2-6q+8}+q-3.
\end{align}
In the thermodynamic limit, the ratio of the site number at the outermost layer to the total lattice site number is $\eta_{n\rightarrow\infty} = (\lambda - 1)/\lambda$. For the $\{4,8\}$ lattice, this ratio is $\eta_{n\rightarrow\infty} = 0.899$.
\begin{figure}[!ht]
\centering
\includegraphics[width=2in]{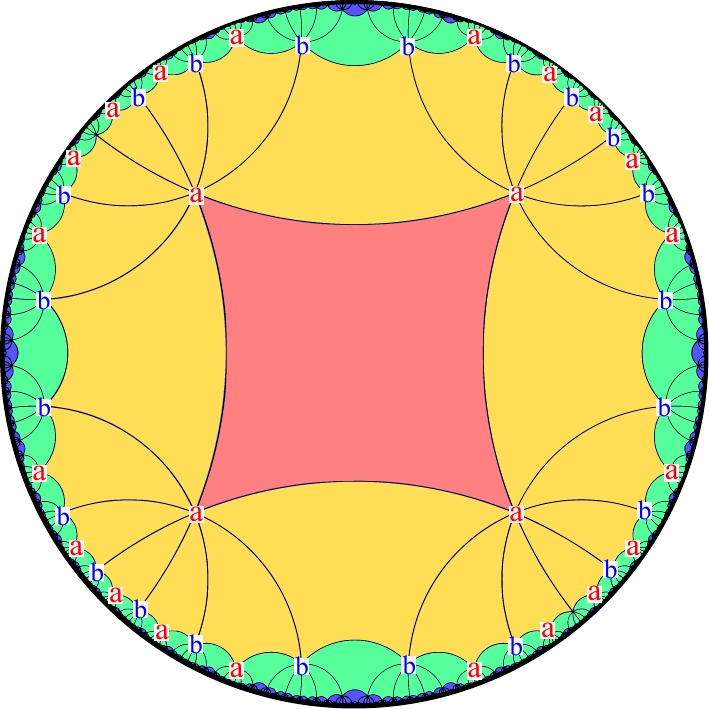}
\caption{{\bf Inflation pattern of the hyperbolic $\{4,8\}$ lattice on the Poincaré disk.} Each epoch (layer) marked in different colors. The symbol $a$ ($b$) labels vertices with two (three) neighboring sites in each epoch.}\label{48}
\end{figure}

\section{Supplementary Note 2: \uppercase{Level statistics under open boundary conditions}}

Achieving the thermodynamic limit on the hyperbolic lattice is a highly nontrivial issue arising from the finite boundary/bulk ratio. If we attempt to reach the thermodynamic using open boundary conditions (OBC), no localization phase transition occurs due to boundary effects. In Supplementary Figure \ref{open}, we present the variation of level statistics $\langle r\rangle$ with respect to the disorder strength for different system sizes of the $\{3,8\}$ lattices (Supplementary Figure \ref{open}(a)) and $\{4,8\}$ lattices (Supplementary Figure \ref{open}(b)) under open boundary conditions. The different curves correspond to various system sizes. As observed, the level statistics $\langle r \rangle$ decrease as the system size increases, and the absence of curve crossings suggests that no phase transition occurs. 

To eliminate the boundary effect, one way is to implement periodic boundary conditions (PBC). To accomplish this, it is necessary to identify pairs of boundary edges within a finite hyperbolic lattice, resulting in a tessellation of a high-genus Riemann surface.  This can be achieved by computing the quotient group of $\Gamma$ (the translation symmetry group of the hyperbolic lattice) and $G$ (a normal subgroup of $\Gamma$), where each element of this quotient group corresponds to a site in the PBC lattices. Moreover, ensuring that the PBC clusters approximate the thermodynamic limit requires constructing a coherent sequence of finite-index normal subgroups \cite{AT2}. 

Instead, our strategy to  eliminate the boundary effect involves connecting all boundary sites randomly.  An advantage of this choice is that its relationship to random regular graphs as $p\rightarrow \infty$ are more apparent in this framework. Additionally, this approach is more feasible for experimental purposes, acknowledging the complexities involved in achieving PBC, which requires meticulous effort. \\

\begin{figure}[!ht]
\centering
\includegraphics[width=4.6in]{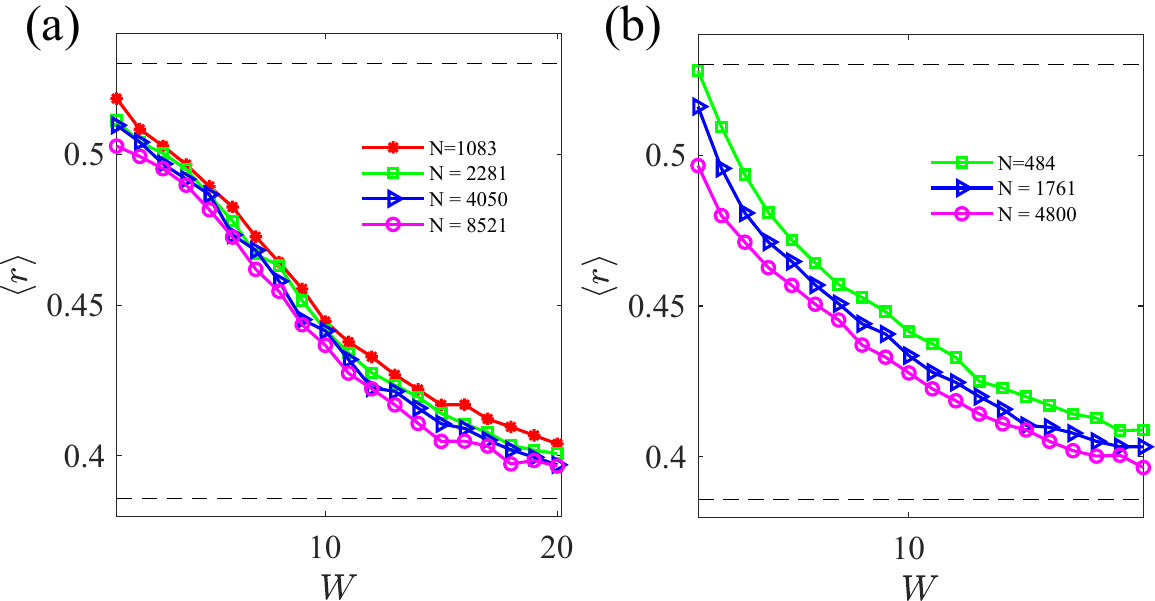}
\caption{{\bf Level statistics with open boundary conditions.} The average ratio of level spacings $\langle r\rangle$ as a function of the disorder strength $W$ with different system sizes for (a) $\{p, q\} = \{3,8\}$ and (b) $\{p, q\} = \{4,8\}$ lattice. The average is taken over the central $1/10$ eigenstates with $60\sim 1000$ disorder realizations. The horizontal lines mark the value $\langle r\rangle=0.387$ and  $\langle r\rangle=0.529$.}\label{open}
\end{figure}

\section{Supplementary Note 3: \uppercase{ The spectrum of hyperbolic Hamiltonian}}
We obtained the spectra numerically by exactly diagonalizing the Hamiltonian matrix. In the real space basis, the Hamiltonian matrix takes the form
\begin{align}
\mathcal{H} = t\mathcal{A} + \mathcal{D},
\end{align}
where $\mathcal{A}$ is the adjacency of $\{p,q\}$ lattices. $\mathcal{D}$ is a diagonal matrix, with diagonal elements $D_{ii} = \epsilon_i$ randomly drawn from the uniform distribution $[-W/2,W/2]$.

A more complicated task is obtaining the adjacency matrix $\mathcal{A}$. It's an $N\times N$ symmetric matrix, and $\mathcal{A}_{ij}$ is nonzero if and only if $i$ and $j$ are connected, $\mathcal{A}_{ij}=\mathcal{A}_{ji}=1$. There are two ways to inflate the hyperbolic lattices, one is inflation start from a polygon, and the other way is inflation from a vertex. \\

When starting from a polygon, where the vertices constitute the set $N_0 = \{S_{0}^{(1)}, S_{0}^{(2)},...,S_{0}^{(p)}\}$, we proceed to obtain the Hamiltonian matrix using the following procedure:  \\
step 1.  We add tiles by attaching $S_{0}^{(1)}$ clockwise (or counterclockwise) until $S_{0}^{(1)}$ are saturate (with $q$ neighbors). During this process, new sites $S_{1}^{(1)},S_{1}^{(2)},...$ are adding. \\
step 2. We then choose another vertex $S_{0}^{(i)}$ from the set $N_0$ sequentially  and repeat step 1. Once all vertices of $N_0$ are saturated, we complete our first layer of inflation. The newly added sites constitute the set $N_1 = \{S_{1}^{(1)}, S_{1}^{(2)},...\}$. \\
step 3. The sites generated in the $(i-1)$-th epoch constitute the set $N_{i-1} = \{S_{i-1}^{(1)}, S_{i-1}^{(2)},...\}$. For the $i$-th epoch, we add tiles by attaching them clockwise to $S_{i-1}^{(1)}$, then repeat steps 1 and 2.\\
step 4. After the $n-th$ epoch, all the sites are saturated except those belonging to $N_n$. Subsequently, we randomly connect the sites in $N_n$ until all sites are saturated. \\
Finnally, we obtain $\mathcal{H}$ and exactly diagonalize it to get the spectrum.\\

 In the case of start from a vertex, we have $N_0 = \{S_{0}\}$, and similary repeat steps 1-4.\\

\section{Supplementary Note 4: \uppercase{Finite-Size scaling analysis methods}}

 We use finite size scaling to get the transition point $W_c$ and the critical exponents $\nu$.  
We define the error function as

\begin{align}
\mathcal{E}_r = \frac{1}{W_s^{max}-W_s^{min}}\int_{W_s^{min}}^{W_s^{max}}dW_s Var_L\{X_L[W_sL^{-1/\nu}+W_c]\}.
\end{align}
where $X_L(W)$ reperesents either $\langle r(W,L)\rangle$ or $\langle IPR(W,L)\rangle^{-1}L^{-\gamma/\nu}$, and $Var_L$ is the variance over different sizes $L$. Since we only have discrete values of $\langle r(W,L)\rangle$ and $\langle IPR(W,L)\rangle$, for each $W_s$, we obtain $X_L[W_sL^{-1/\nu}+W_c]$ by interpolation. By choosing appropriate $W_s^{min}$ and $W_s^{max}$, we determine $W_c$ and $\nu$ (or $\gamma$) by minimizing $\mathcal{E}_r$.  If we choose $W_s^{min}$ and $W_s^{max}$ from a box distribution, such as $W_s^{min}\in [-200,-100]$ and  $W_s^{max}\in [100,200]$, we obtain different values of $W_c$ and $\nu$ (or $\gamma$), providing an error estimate of $W_c$ and $\nu$ (or $\gamma$).

\section{Supplementary Note 5: \uppercase{ Finite-size scaling analysis of AT at band bottom}}
One notable characteristic of the presence of mobility edges is the distinct critical transition points of Anderson localization for eigenstates within different segments of the energy spectra. In the main text, we identified the transition point at $W_c = 77.2$ through finite-size scaling of $\langle r\rangle$ (average ratio of level spacings) and at $W_c = 83$ through finite-size scaling of $\langle IPR\rangle$ (average inverse participation ratio). In this section, we perform finite-size scaling for eigenstates at the band bottom (1/10 portion) of the hyperbolic $\{3,8\}$ lattice. Our objective is to determine the transition point and critical exponents associated with this spectral region.

\begin{figure}[!ht]
\centering
\includegraphics[width=4.4in]{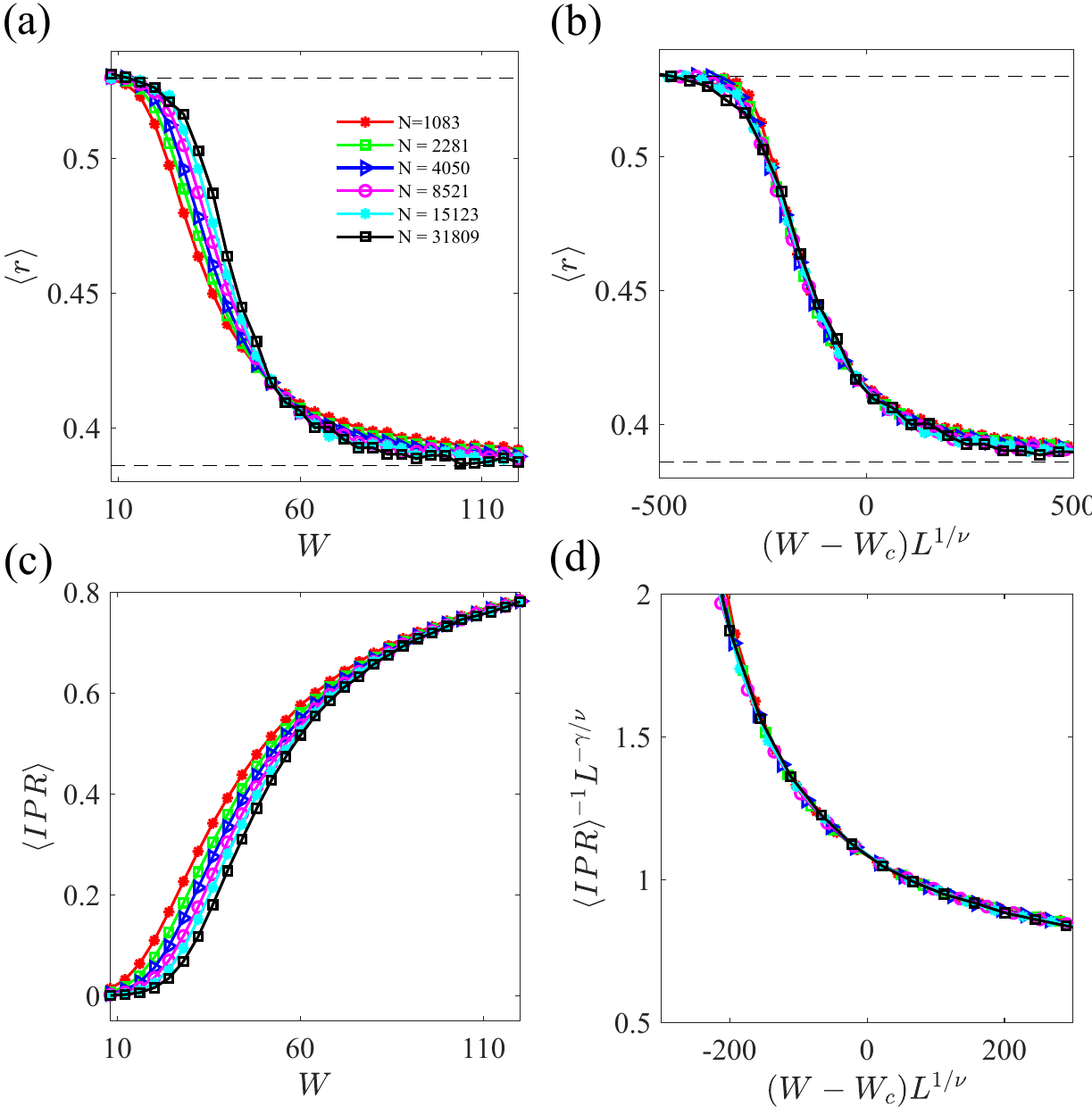}
\caption{{\bf Finite-size scaling analysis of Anderson transition (AT) at the band bottom of hyperbolic $\{3,8\}$ lattices}. (a) (b) Average ratio of level spacings, $\langle r\rangle$, as a function of disorder strength for various system sizes before and after scaling, with fitting function $\langle r\rangle=f[(W-W_c)L^{1/\nu}]$. (c)(d) Average IPR, $\langle IPR\rangle$, as a function of disorder strength for different system sizes before and after scaling, with fitting function $\langle IPR\rangle^{-1}L^{-\gamma/\nu}=g[(W-W_c)L^{1/\nu}]$. The averages are taken over $20\sim 1000$ samples and the lowest $1/10$ of the eigenstates. The horizontal lines in (a) and (b) mark the values $\langle r\rangle=0.387$ and $\langle r\rangle=0.529$. In (b), the best fit yields the transition point $W_c =54.4\pm 5$ and $\nu = 0.97\pm 0.05$. In (d), the best fit yields the transition point $W_c =62\pm 5$ and $\gamma = 0.23\pm 0.05$.}\label{figS1}
\end{figure}

Supplementary Figure \ref{figS1}(a) plots the variation of $\langle r\rangle$ with respect to the disorder strength for different system sizes, reaching up to $N=31809$. The average is computed over $20\sim 1000$ disorder samples, and the lowest $1/10$ of the eigenenergies. Note that for larger system sizes, the computation is time-consuming, thus fewer samples are taken. The value descends from $\langle r\rangle=0.529$ (delocalized region) at small disorder strength to $\langle r\rangle=0.387$ (localized region) at larger disorder strength. To identify the AT, we employ the fitting function $\langle r\rangle=f[(W-W_c)L^{1/\nu}]$, where $L={\rm ln}N$, $W_c$ is the transition point, and $\nu$ denotes the associated critical exponent. Our numeric suggests the best fit to be $W_c = 54.4\pm5$ and $\nu = 0.97\pm0.05$, as shown in Supplementary Figure \ref{figS1}(b). For the IPR, we observe its change from $0$ to a finite value with increasing disorder strength, as depicted in Supplementary Figure \ref{figS1}(c). Using the fitting function $\langle IPR\rangle^{-1}L^{-\gamma/\nu}=g[(W-W_c)L^{1/\nu}]$, where $\nu$ is the same critical exponent obtained from the scaling of $\langle r\rangle$, and $\gamma$ is another critical exponent, our numeric suggests an optimal fit of IPR when $W_c = 62.0\pm5$ and $\gamma = 0.23\pm 0.03$. As illustrated in Supplementary Figure \ref{figS1}(d), the IPR curves associated with different system sizes coincide under this scaling. To summarize, there is a noticeable discrepancy in the critical disorder strength of AT at the band center and band bottom, signifying the presence of mobility edges.

\section{Supplementary Note 6: \uppercase{ Fractal dimensions}}

 It is well established that the eigenstates of disordered systems near the transition point exhibit fractal behavior, characterized by specific fractal dimensions \cite{hoffmann, schreiber}. Consider the $q^{th}$ moment of the wave function,

\begin{align}
M_q = \sum_{n=1}^N |\phi_n|^{2q}
\end{align}

where $\phi_n$ is the $n$-th component of the eigenfunction $|\phi\rangle$ around the transition point. Notably, $M_2$ corresponds to the inverse participation ratio (IPR). Near the transition point, $M_q$ scales as $\langle M_q\rangle \sim L^{-\tau_q}$, where the average is taken over all eigenstates and disorder configurations. The fractal dimension is then defined by $D_q = \tau_q / (q-1)$.

Focusing on $q=2$, we have $\langle M_2\rangle = \langle IPR\rangle$. We show the  $\ln(\langle IPR \rangle)$ versus $\ln (L)$ of the  $\{3,8\}$ ($\{4,8\}$) lattice around the transition point $W \approx 80 (92)$ in Supplementary Figure \ref{fractal_D}. The slope of which corresponds to fractal dimensions $D_2^{\{3,8\}} = 0.3\pm 0.02$  and  $D_2^{\{4,8\}} = 0.28\pm0.03$.

\begin{figure}[!ht]
\centering
\includegraphics[width=4.6in]{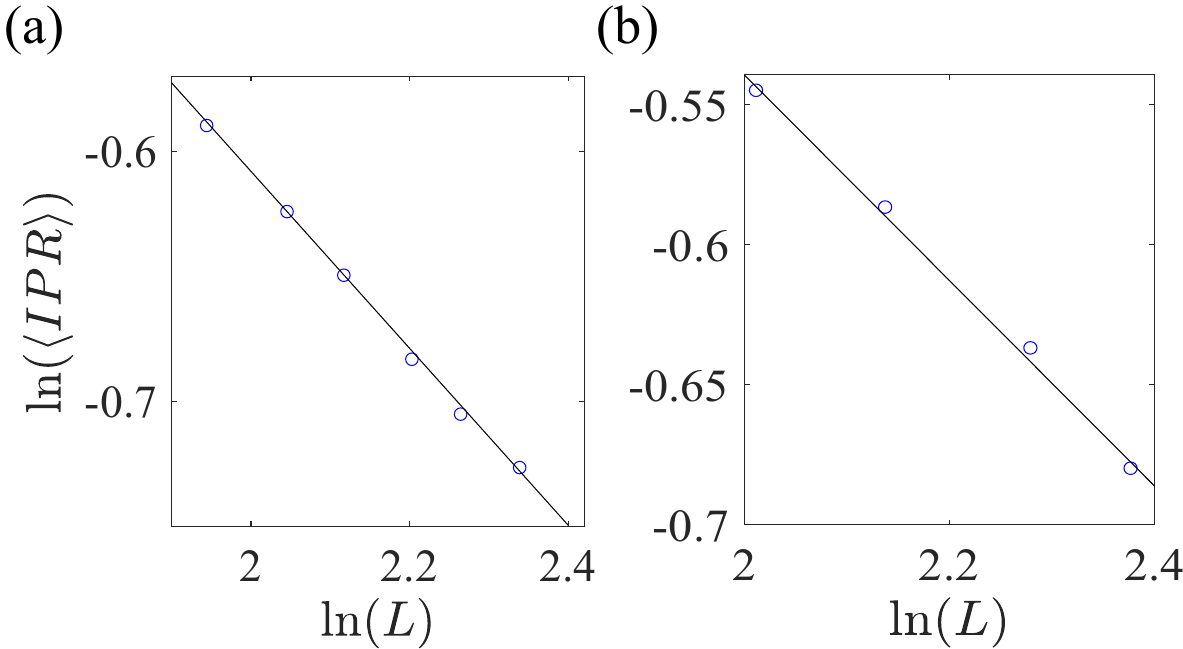}
\caption{{\bf Fractal dimensions.} The $\ln(\langle IPR \rangle)$ versus $\ln (L)$ of the  $\{3,8\}$ lattice (a) and $\{4,8\}$ lattice (b) at the transition point. The slope of which corresponds to $D_2^{\{3,8\}} = 0.3\pm 0.02$  and  $D_2^{\{4,8\}} = 0.28\pm0.03$. We take the transition point $W_c \approx 80 (92)$ for $\{3,8\}$ ( $\{4,8\}$) lattices. }\label{fractal_D}
\end{figure}

\section{Supplementary Note 7: \uppercase{ Two-sided scaling analysis}}
Instead of the scaling analysis performed in the main text, we consider another scaling method used in Refs. \cite{AT2,prb2022}, which is performed from two sides of the AT. We focus on two quantities, the average ratio of level spacings, $\langle r\rangle$, and $\langle IPR\rangle$. The scaling function has the form:
\begin{align}
X(W,L) = X(W_c,L)F(L/\xi(W)).
\end{align}
Here, $X$ can be either $\langle r\rangle$ or $\langle IPR\rangle$, and $F$ is the scaling function that depends on the ratio of system size $L$ and the characteristic length $\xi$. The latter depends on the disorder strength $W$ and diverges at the transition point $W_c$ as $\xi \propto |W-W_c|^{-\alpha}$, where $\alpha$ is the critical exponent. We use the same data as in Fig. 3 and Fig. 4 of the main text and organize this data $X(W,L)$ into sets $\{X_{W_i}(L)\}_{i=1}^{M+1}$ based on the disorder strength $W$. We further assume a transition point $W_c$ and normalize all data sets by the critical data, denoted as $\{\hat X_{W_i}(L)\}_{i=1}^{M}$, with $\hat X_{W_i}(L) = X_{W_i}(L)/ X_{W_c}(L)$. The scaling procedures are detained below.

For the delocalized side, assuming there exists $S$ data sets that satisfy $W < W_c$. We initiate the process with the first dataset, $\hat X_{W_1}(L)$, located furthest from $W_c$. Subsequently, we rescale the $x$ axis by $L/\xi(W_2)$ for the second dataset, $\hat X_{W_2}(L)$. The characteristic length, $\xi(W_2)$, is chosen to optimize the collapse of the rescaled second dataset, $\hat X_{W_2}(L/\xi(W_2))$, onto $\hat X_{W_1}(L)$, with the quality of the collapse assessed using a $\chi^2$ test. To perform the $\chi^2$ test, linear interpolation is employed to obtain the value of $\hat X_{W_1}(L/\xi(W_2))$; any values of $L/\xi(W_2)$ falling outside the range of the first dataset are disregarded. The $\chi_{W_2}^2$ index is then computed as $\sum_i (\hat X_{W_1}(L/\xi(W_2))-\hat X_{W_2}(L/\xi(W_2)))^2$. We repeat this procedure by rescaling the $x$ axis with $L/\xi(W_3)$ for the third dataset, $\hat X_{W_3}(L)$, ensuring a collapse onto $\hat X_{W_2}(L/\xi(W_2))$, and record the values of $\xi(W_3)$ and $\chi_{W_3}^2$. This process iterates, yielding the values $\xi_{W_i}$ and $\chi_{W_i}^2$ for subsequent iterations ($i = 2,3,...,S$).
\begin{figure}[!ht]
\centering
\includegraphics[width=4.6in]{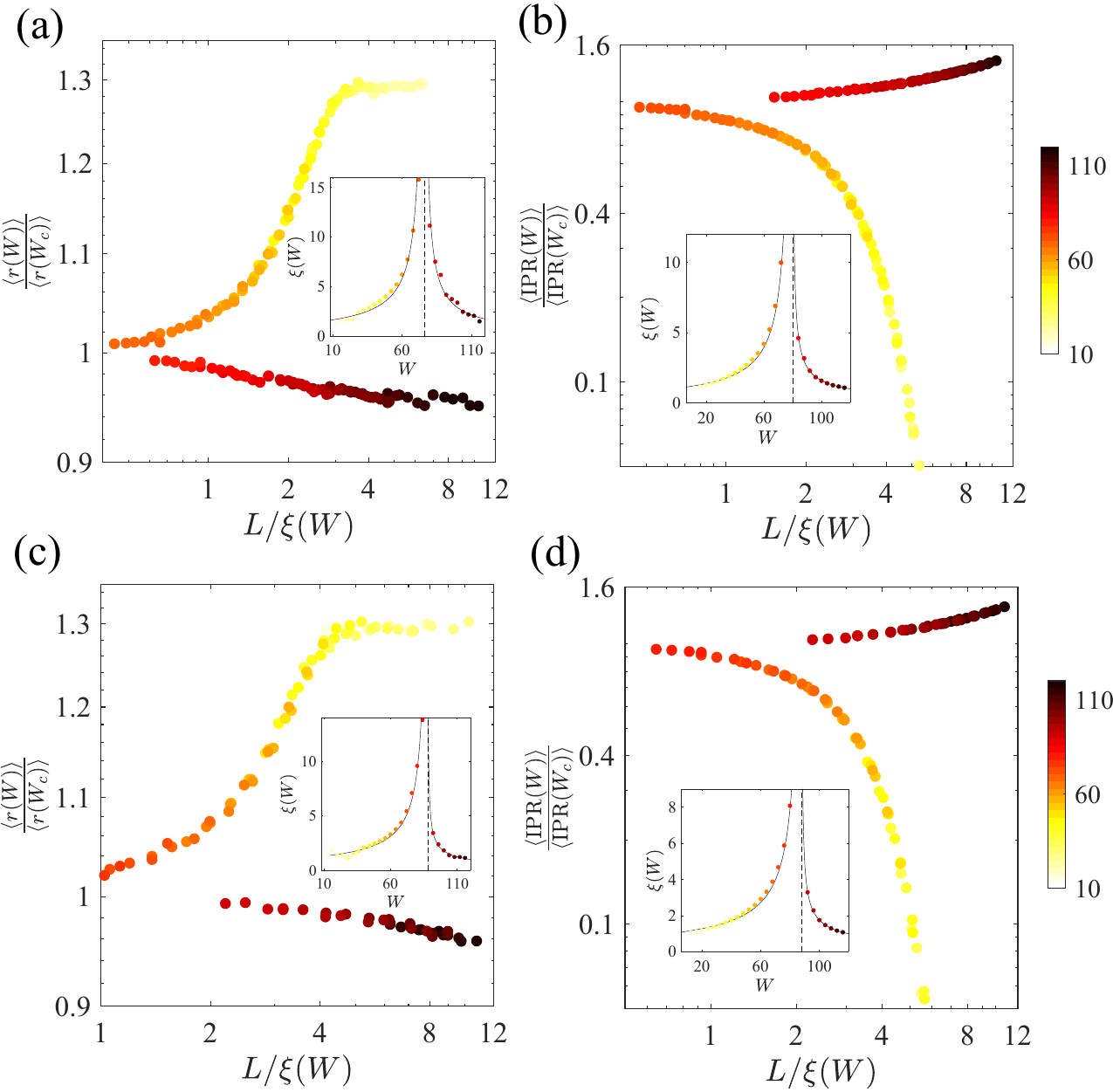}
\caption{{\bf Two-sided scaling analysis.} The two-sided scaling analysis of the AT on (a,b) hyperbolic $\{3,8\}$ and (c,d) $\{4,8\}$ lattices. (a,c) The scaling of the quantity $\frac{\langle r(W)\rangle}{\langle r(W_c)\rangle}$ from two sides of the AT. (b,d) The scaling of $\langle IPR\rangle$ (average IPR) from two sides of the AT. (Insets) Near the transition point $W_c$, the characteristic length diverges as $\xi(W) \propto |W-W_c|^{-\alpha}$, with $\alpha$ the critical exponent. The transition point extracted from the $\chi^2$-test is taken as $W_c = 80$ for the $\{3,8\}$ lattice in (a,b) and $W_c = 88$ for the $\{4,8\}$ lattice in (c,d). The color bar denotes the disorder strength.}\label{figS2}
\end{figure}

For the localized side, choose the data sets that satisfy $W > W_c$. Begin with the last dataset, $\hat X_{W_M+1}(L)$ that is farthest from $W_c$. Following the aforementioned procedure, obtain values of $\xi_{W_i}$ and $\chi_{W_i}^2$ for iterations $i = S+2, S+3, ..., M$. The total sum of $\chi^2(W_c)$ across both delocalized and localized sides is then employed to pinpoint $W_c$, which minimizes the summation and is identified as the transition point. The two-sided scaling of the two quantities $\langle r(W,L)\rangle$ and $\langle {\rm IPR}(W,L)\rangle$, on $\{3,8\}$ lattice [Supplementary Figure \ref{figS2}(a)(b)] and $\{4,8\}$ lattice [Supplementary Figure \ref{figS2}(c)(d)] is illustrated in Supplementary Figure \ref{figS2}. Through the above scaling procedure, the original data [shown in Fig. 3(a)(c) and Fig. 4(a)(c) of the main text] collapses into a single curve. From the $\chi^2$-value analysis, the critical disorder strength $W_c$ minimizing $\chi^2(W_c)$ for $\{3,8\}$ lattice is $W_c = 80\pm 4$, with critical exponents extracted from the characteristic length  $\{\alpha_{\langle r\rangle,{\rm del}}^{\{3,8\}},\alpha_{\langle r \rangle,{\rm loc}}^{\{3,8\}}\}=\{0.89\pm 0.11,0.85\pm0.14\}$ and $\{\alpha_{\langle{\rm IPR}\rangle,{\rm del}}^{\{3,8\}},\alpha_{\langle {\rm IPR}\rangle,{\rm loc}}^{\{3,8\}}\}=\{0.97\pm 0.07,0.69\pm 0.04\}$. For $\{4,8\}$ lattice, the minimum $\chi^2(W_c)$ occurs at $W_c = 88\pm 4$, with critical exponents $\{\alpha_{\langle r\rangle,{\rm del}}^{\{4,8\}},\alpha_{\langle r \rangle,{\rm loc}}^{\{4,8\}}\}=\{0.87\pm 0.1,0.60\pm 0.1\}$ and $\{\alpha_{\langle{\rm IPR}\rangle,{\rm del}}^{\{4,8\}},\alpha_{\langle {\rm IPR}\rangle,{\rm loc}}^{\{4,8\}}\}=\{0.86\pm 0.05,0.59\pm 0.04\}$. This is consistent with the critical disorder strengths presented in the main text [$W_c=83\pm 4$ for the $\{3,8\}$ lattice and $W_c=94\pm 4$ for the $\{4,8\}$ lattice], as determined by another scaling method.

\end{document}